\documentclass{article}
\usepackage{graphicx} 
\usepackage{caption}
\usepackage{authblk}
\usepackage{hyperref}
\usepackage{amsmath}
\usepackage{amsfonts}
\usepackage{csquotes}
\usepackage[top=1in,left=1.35in, right=1.35in]{geometry}

\usepackage{amssymb, mathtools, bbm}
\mathtoolsset{showonlyrefs=true}

\title{Integrated Information in the Active Inference Framework}
\author[1,2]{Carlotta Langer} 
\author[3]{Jesse van Oostrum} 
\author[3,4,5]{Nihat Ay}
\affil[1]{Max Planck Institute of Molecular Cell Biology and Genetics, Dresden, Germany}
\affil[2]{Center for Systems Biology Dresden, Dresden, Germany }
\affil[3]{Hamburg University of Technology, Hamburg, Germany}
\affil[4]{Santa Fe Institute, Santa Fe, USA}
\affil[5]{Leipzig University, Leipzig, Germany}
\date{November 2024}

\usepackage{parskip}
\usepackage{mymacros}
\newcommand{\DKL}{D_{\mathrm{KL}}}
\usepackage{tikz}
\usetikzlibrary{positioning, calc}
\tikzstyle{neuron} = [circle,draw, minimum size=.9cm, inner sep=0]

\usepackage{xcolor}
\usepackage[backend=bibtex ,natbib=true,giveninits=true,maxnames=25]{biblatex}
\begin{document}
\maketitle

\begin{abstract}
The active inference framework provides a principled approach to modeling sentient behavior. In this framework perception and action selection are treated
in a unified way. The resulting agents form an internal generative model of the
relevant dynamics of the world in order to infer their future observations, their
internal states and to select actions. We combine this modeling framework with
the Integrated Information Theory of consciousness and are therefore able to
analyze the active inference agents from the perspective of integrated
information. The Integrated Information Theory aims at quantifying the level of
consciousness of a system by assessing its capability to integrate information.
Here, we define a measure of integrated information for the generative
model by making an additional structural assumption. Experiments with
simulated agents reveal a correlation between integrated information measures and the free energy of the active inference
agents that increases with the size of the generative model. 
\end{abstract}

\section{Introduction}

The active inference framework aims to model perception and action selection by minimizing \enquote{free energy} of a generative model. This internal model describes the environmental dynamics that are relevant for the agent and is used for perception and action selection. In the sketch on the left of Figure \ref{Fig:1} this is illustrated by a cat's mental image of the moving mouse toy.  Hence, this theory relies on the agent forming an internal model of the dynamics of the environment.

On the other hand, the Integrated Information Theory does not consider the relationship of the processes in the brain, or controller, to the environment. Instead, it focuses solely on the connections in the brain, as sketched in Figure \ref{Fig:1} on the right. This theory aims at quantifying consciousness by assessing the amount of information integration in the brain. In this case the interaction of the agent with the environment, as well as its actuator or sensory signals are not of interest. 

It has been suggested that these two influential neuro-scientific theories might be compatible and that a minimization of free energy could lead to a maximization of integrated information, as discussed in \cite{FristonWiese2020, Olesen}. 
In \cite{CORCORAN2026106742} the authors sketch similarities and differences of these theories that lead to diverging predictions regarding human and animal consciousness, which are being tested in an ongoing adversarial collaboration. 

Here, we slightly adapt the generative model of active inference agents in order to calculate various integrated information measures to quantify the information flow inside the generative model. We then train simulated agents via learning rules that have been suggested in the active inference literature and observe that minimizing the free energy is correlated to some of the integrated information measures. 

\begin{center}
    \includegraphics[height = 0.15\textheight]{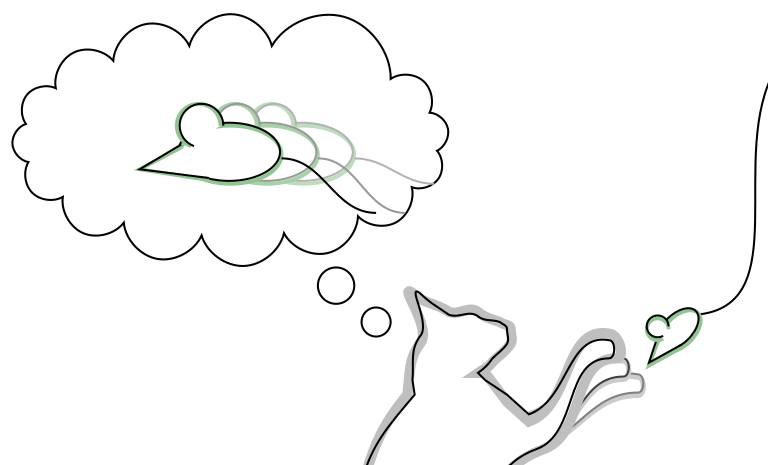} \hspace{2cm}
        \includegraphics[height = 0.15\textheight]{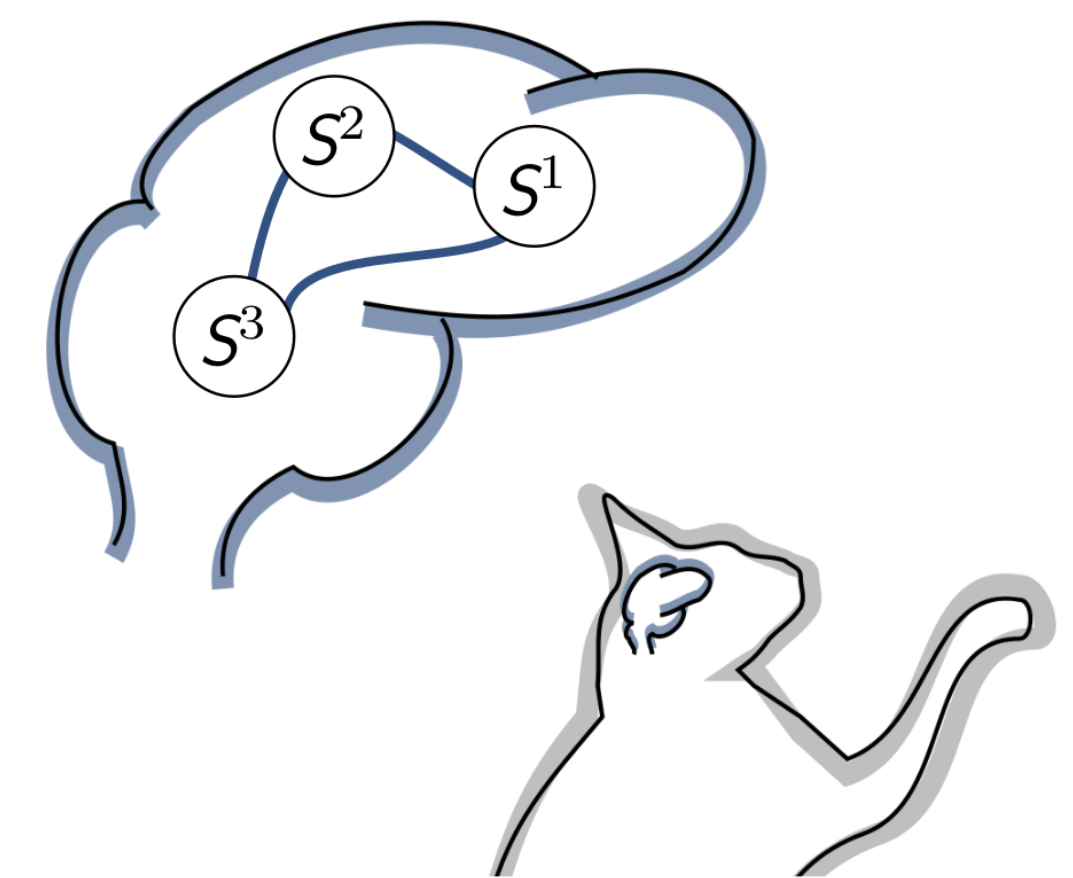} 
        \captionof{figure}{Sketches illustrating the generative model in the active inference framework that aims at describing world dynamics on the left. The sketch on the right illustrates the integrated information theory in which the structure of the connections in the brain are of interest.} \label{Fig:1}
\end{center}

In the following Section \ref{sect:relatedwork} we first give a brief introduction to related work. Then we define the methods of the theories of active inference and integrated information in more detail in Sections \ref{sect:activeinference} and \ref{sect:IIT} and combine both theories in one approach in Section \ref{sect:combining}. Afterwards, we introduce our experiment in Section \ref{sect:setting} and discuss the results in Section \ref{sect:results}. 

The experiments were programmed in python 3 and the code and data of our experiments are available at \cite{Langer_git_2026}.

\subsection{Related Work} \label{sect:relatedwork}

In this work, we connect the active inference framework with the integrated information theory of consciousness. Both theories apply to systems that are modeled via stochastic processes and make use of information-theoretic quantities such as suprisal, entropy or KL-divergence. Information theory is based on Shannon's mathematical theory of communication and provides a framework to analyze the properties of a communication channel, \cite{Shannon}. Since both theories, active inference and IIT, can be applied to similar systems and phrased in the common language provided by information theory, we are able to combine them and analyze their relationship. 

The active inference framework emerged from the free energy principle (FEP), which was originally introduced by Friston \cite{friston2010free} as a unifying account of perception, action, and learning in biological systems. The core idea of the FEP is that any self-organizing system that maintains its integrity over time \cite{kirchhoff2018markov} must minimize an upper bound on the surprisal of its sensory states, namely variational free energy. Active inference extends this principle by treating action as an inferential process: actions are selected so as to fulfill predictions of an internal generative model, thereby closing the perception--action loop.

A key motivation of active inference lies in embodied and enactive theories of cognition, where cognition is understood as arising from the ongoing interaction between an agent and its environment \cite{clark1997being}. Rather than passively representing the world, the agent actively samples its environment in ways that are consistent with its prior beliefs and preferences. Perception corresponds to Bayesian inference over hidden states of the world, while action selection corresponds to inference over policies that are expected to minimize future free energy. This formulation allows perception, learning, and action to be expressed within a single probabilistic framework \cite{parr2022active}.

Active inference has been developed in both continuous and discrete state-space settings. In particular, discrete-state formulations based on partially observable Markov decision processes (POMDPs) have become widely used due to their conceptual clarity and computational tractability. In this setting, the generative model is specified by conditional probability distributions encoding state transitions, observation likelihoods, and prior preferences over outcomes. Policy selection is then implemented by minimizing expected free energy, which balances epistemic value (information gain) and pragmatic value (goal fulfillment), as discussed in \cite{DaCosta} and \cite{concise}. These discrete formulations provide a natural setting for simulations and learning experiments, and form the basis for the agents studied in the present work.

The Integrated Information Theory (IIT) of consciousness was introduced mainly by Giulio Tononi and his coworkers and it aims at assessing the amount and quality of consciousness of a system. This theory went through various phases of development. It started as a measure of brain complexity \cite{Tononi1994} and developed into an involved theory of consciousness through IIT 1.0  \cite{Tononi2003, Tononi2004}, IIT 2.0 \cite{Tononi2008}, IIT 3.0 \cite{Oizumi2014} and, most recently, IIT 4.0 \cite{IIT40}. The main idea, namely measuring how much a system differs from one that is not capable of integrating information, stays consistent throughout the different versions of the theory. The approach of the theory is to start from phenomenology translated into axioms about consciousness and then deduce postulates and a mathematical formalism from these axioms. Hence, the calculation of the measures changes significantly, from the mathematical perspective, while the axiomatic definition remains in large parts unchanged. In this work, we calculate one measure that can be seen  in the context of IIT 2.0, \cite{Langer2020}, one measure for IIT 3.0 and two measures in the context of IIT 4.0. 

The IIT measures focus solely on the internal states of the agents, meaning the states of the \enquote{brain} or \enquote{controller}.
Nevertheless, some experiments with embodied agents and integrated information have been conducted. The authors of \cite{Albantakis2014} and \cite{Albantakis2015} analyze the integrated information in evolving and adapting simulated agents. Furthermore, in \cite{Edlund} the authors conclude that the integrated information increases with the fitness of simulated agents. 
In \cite{Langer2021} we analyze the relationship between the integrated information and the interaction of the agent with the environment in acting, simulated agents and in \cite{Langer2024} we observe this relationship in learning agents.

It has been suggested that these two influential neuro-scientific theories, integrated information and active inference, might be compatible and that a minimization of free energy could lead to a maximization of integrated information, see  \cite{FristonWiese2020}.  Furthermore, the authors of \cite{Olesen} analyze the integrated information value of animats and observe that integrated information fluctuates with the surprisal of the system. Here, we analyze simple artificial agents that learn to interact with their environment using the active inference framework and we compare the various integrated information measures with the free energy. We observe that a smaller value of the free energy is indeed, weakly in some cases, correlated to a higher integrated information value and that this correlation strengthens with the size of the generative model. 

\section{Methodology} \label{sect:methodology}

In this paper we consider an active inference agent acting in a discrete-time setting with a finite-time horizon. This means that we consider a sequence of $T$ time steps and at every time step $\tau$ the agent receives an observation $o_\tau$, and performs an action $a_\tau$. We use $\tau$ for arbitrary time steps and the letter $t$ to denote the current time step. We use the subscript ${}_{\tau:\tau'}$ to denote a sequence of random variables and outcomes, e.g.\ $o_{\tau:\tau'} = (o_\tau, \ldots, o_{\tau'})$. We denote random variables with capital letters and it's outcomes with small letters, e.g. $X$ and $x$ respectively. We denote a probability distribution of a random variable $X$ by $p(X)$ and its entries by $p(x)$, shorthand for $p(X=x)$.
 

\subsection{Active Inference} \label{sect:activeinference}

\begin{figure} [ht]
    \centering
    
    \begin{tikzpicture}
\draw[->, line width = 0.3mm] (0,0)--(1.5,0);

\draw[->, line width = 0.3mm] (2,0)--(3.5,0);

\draw[->, dotted, line width = 0.3mm] (4,0)--(6,0);

\draw[fill = white] (0.75,1.5) circle (0.365cm);
\draw[fill = white] (2.75,1.5) circle (0.365cm); 

\draw[->, line width = 0.3mm] (0,0)--(0,-1);
\draw[->, line width = 0.3mm] (2,0)--(2,-1);
\draw[->, line width = 0.3mm] (4,0)--(4,-1);
\draw[->, line width = 0.3mm] (6.5,0)--(6.5,-1);
\draw[fill = white] (0,-1.5) circle (0.4cm);
\draw[fill = white] (2,-1.5) circle (0.4cm);
\draw[fill = white] (4,-1.5) circle (0.4cm);
\draw[fill = white] (6.5,-1.5) circle (0.4cm);
\draw[fill = white] (0,0) circle (0.4cm);
\draw[fill = white] (2,0) circle (0.4cm);
\draw[fill = white] (4,0) circle (0.4cm);
\draw[fill = white] (6.5,0) circle (0.4cm);
\draw[->, line width = 0.3mm] (0.75,1.5)--(1.7,0.45);
\draw[->, line width = 0.3mm] (2.75,1.5)--(3.7,0.45);
\draw[->, line width = 0.3mm] (5.25,1.5)--(6.2,0.45);
\draw[fill = white] (0.75,1.5) circle (0.4cm);
\draw[fill = white] (2.75,1.5) circle (0.4cm);
\draw[fill = white] (5.25,1.5) circle (0.4cm);

\draw[] (0,0) node{$S_{1}$};
\draw[] (2,0) node{$S_{2}$};
\draw[] (4,0) node{$S_{3}$};
\draw[] (6.5,0) node{$S_{T}$};
\draw[] (0,-1.5) node{$O_{1}$};
\draw[] (2,-1.5) node{$O_{2}$};
\draw[] (4,-1.5) node{$O_{3}$};
\draw[] (6.5,-1.5) node{$O_{T}$};

\draw[] (0.75,1.5) node{$a_{1}$};
\draw[] (2.75,1.5) node{$a_{2}$};
\draw[] (5.25,1.5) node{$a_{\scriptscriptstyle T-1}$};
\end{tikzpicture}
    \caption{Graphical representation of the generative model}
    \label{fig:representation-genmod}
\end{figure}
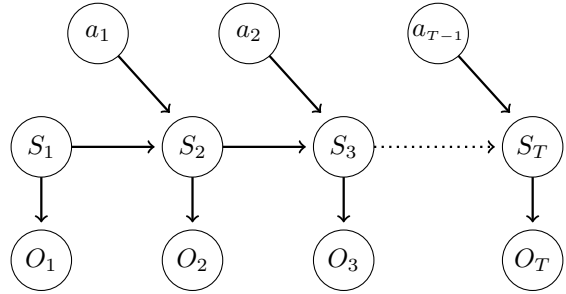

The agent models the dynamics of the environment using an internal generative model. This model uses a variable $S_\tau$, called an internal state, to represent the state of the environment at the time step $\tau$. The model is given by the following probability distribution:
\begin{align}
    p(O_{1:T}, S_{1:T} | a_{1:T-1}, \theta),
\end{align}
where $\theta$ is used to parametrize the model according to the graph in Figure \ref{fig:representation-genmod}. We will sometimes suppress the dependence on $\theta$ for ease of notation. Algebraically, the distribution factorizes as follows:
\begin{align}
p(O_{1:T}, S_{1:T} | a_{1:T-1}) &= p(S_1) p(O_1 \mid S_1) 
 \prod_{\tau=2}^{T} p(S_\tau \mid S_{\tau-1}, a_{\tau-1}) p(O_\tau \mid S_\tau).
\end{align}

Suppose the agent has received observations $o_{1:t-1}$ and performed actions $a_{1:t-1}$, it can compute its belief about the next observation $O_t$ on the basis of the generative model. This belief is given by a distribution $q$ which is the approximate posterior distribution of the generative model. That is, 
\begin{align}
    q(O_t | o_{1:t-1}, a_{1:t-1}) \approx p(O_t | o_{1:t-1}, a_{1:t-1}, \theta).
\end{align}

In order to quantify how well the agent performs active inference we use the variational free energy (VFE) as a measure for this. The VFE for observation $o_t$ is given by 
\begin{align}
    - \log q(o_t | o_{1:t-1}, a_{1:t-1}) \label{VFE}
\end{align}
where we have now substituted the actually observed value $o_t$ into the belief  $q(O_t | o_{1:t-1}, a_{1:t-1})$ distribution over possible observations. 

For more details on how the agent performs action selection, perception, and learns the generative model, we refer the reader to Appendix \ref{app:act-inf}.

\subsection{Integrated Information Theory} \label{sect:IIT}

Multiple measures have been proposed to calculate the integrated information. In this work we make use of and compare four of these measures. Although their calculation differs significantly between versions, the core ideas behind the measures remain the same. The integrated information measures aim to assess the difference between a system in which information is integrated compared to one that is cut into parts. In \cite{Tononi2008} Tononi summarizes this idea in the following way: \enquote{In short, integrated information captures the information
generated by causal interactions in the whole, over and
above the information generated by the parts.}. The versions of the theory differ mainly in the definition of the applied \enquote{difference} and the right \enquote{partial system}. 

The first measure that we consider was proposed by us in \cite{Langer2020} as ground truth integrated information and it can be seen in the context of Integrated Information 2.0. There we minimize the KL-divergence between the full system $\tilde{p}$ and a split system that consists of all those distributions that describe a system without integrated information. Let $X_t = (X^1_t, ... , X^n_t)$ a random variable corresponding to the system that we want to analyze. Let additionally $W_t$ be a random variable encoding environmental influences on $X_{t+1}$ and let $(X_t,W_t)_{t \in \mathbb{N}}$ be a Markov process with the transition probability $p(x_{t+1}, w_{t+1} | x_t, w_t) = \prod\limits_{i=1}^n p(x^i_{t+1} | x_t, w_t) p(w_{t+1}| w_t)$. Then we define the split system as the system in which there is no influence from any $X^i_t$ on $X^j_{t+1}$ if $i \neq j$. This results in the following set 
\begin{equation*}
    \mathcal{M}_T = \left\lbrace p \text{ prob. distr. } \mid p(x_{t+1}, w_{t+1} | x_t, w_t) = \prod\limits_{i =1}^n p(x^i_{t+1} | x_t^i) \right\rbrace
\end{equation*}
and the measure is a sum of conditional mutual information terms:
\begin{align}
   \Phi_T(X) = \inf\limits_{p \in \mathcal{M}_T}  D_{KL} (\tilde{p} \parallel p) = \sum\limits_{i = 1}^n I(X^i_{t+1} ; X^{I \setminus i}_t | X_t^i, W_t). \label{Phi_t}
\end{align}

The integrated information measure in IIT 3.0 and 4.0, denoted by $\Phi_{3.0}$ and $\Phi_{4.0}$, respectively, follow the same core idea, namely calculating the difference between the full system and a partial one. 
However, the definition of a partial system and the calculation of the difference is much more involved in this case, hence we will only give a rough overview over the calculation here. 

The calculation in the case of IIT 3.0 takes the past and future steps of the system $X$ into consideration and depends on the current state of the system, $x_t$. Here, we do not consider outside influences. A partial system in this case is defined as the unidirectional bipartition that makes the least difference to the causal structure and it is called a \enquote{minimum information partition} $p^{\text{MIP}}$. The calculation of this partition is quite involved and uses the Wasserstein-distance, also known as earth-movers distance. Those subsets of a system that have a maximally irreducible causal role within the system are called \enquote{concepts}, where $\mathcal{C}$ denotes the set of  concepts of a system. Now, $\Phi_{3.0}$ is defined as the extended  Wasserstein-distance distances between the full collection of concepts of a system and their minimum information partitions:
\begin{align}
    \Phi_{3.0}(X \vert x_t) =  D_{EW}(\mathcal{C} \parallel \mathcal{C}^{\text{MIP}}),
\end{align}
which leads to a sum over the concepts in $\mathcal{C}$ of weighted Wasserstein-distances. Note that the distances are not calculated on the distributions underlying the concepts directly, but on the integrated information associated to these concepts, called $\varphi^{\text{Max}}$ values. 
More details regarding the exact calculation of $\Phi_{3.0}$ can be found in \cite{Oizumi2014,krohn_computing_2017} and the mathematical structure is discussed in more detail in \cite{Kleiner}. 

There are many differences between the IIT 3.0 and 4.0 framework, detailed in the supporting information of \cite{IIT40}. The perhaps most notable change is the introduction of a new difference measure, called the \enquote{intrinsic difference} \cite{intrinsicdifference}, which is consistent with IIT's postulates. Let $p$ and $q$ be two probability distributions on the state space $\mathcal{X}$, then the intrinsic difference is given by
\begin{align*}
    D_{ID}(p \parallel q) = \max\limits_{x \in \mathcal{X}} \left\lbrace p(x) \log \left( \dfrac{p(x)}{q(x)} \right) \right\rbrace.
\end{align*}
The subsets of a system that have a causal impact on themselves are now called \enquote{distinctions} instead of concepts and \enquote{relations} specify how the overlapping nodes of different distinctions are connected. The distinctions and relations each have their own way to be quantified via partitioning and the intrinsic difference, leading to $\varphi_d$ and $\varphi_r$. Let $D$ be the set of distinctions of a system and $R(D)$ be the corresponding relations, then \enquote{big phi} is calculated as:
\begin{align}
    \Phi_{4.0}(X \vert x_t) = \sum\limits_{i \in D \cup R(D)} \varphi_i.
\end{align}
    Additionally, we compute a quantity called \enquote{system phi}, denoted by $\phi_{4.0}^s$, which quantifies the integrated information in a system in itself as one unit. This is defined using the intrinsic difference between the full system and the minimum information partition. More details about the computation of these measures can be found in \cite{intrinsicdifference, IIT40, systemphi}.


Implementations of the IIT 3.0 and 4.0 measures can be found at \cite{PyPhi}.

\subsection{Calculating Integrated Information in an Active  Inference Agent} \label{sect:combining}

In this section, we describe the additional structure that we impose on the generative model of the active inference agent to calculate different integrated information measures. 

The core idea of the integrated information measures is to calculate the difference that integrating information among the parts of the system makes for the system internally, as discussed in the previous section. 
Hence, we need to differentiate between different internal parts of the main object of active inference: the generative model. There we now divide the random variable corresponding to the internal node $S_t$ into a vector of random variables 
\begin{align}
    S_t = (S_t^1, ... , S_t^n),
\end{align}
such that the variables $S_t^i, S_t^j$ are independent given $S_{t-1}$, with $i,j \in \{1,\ldots,n\}, i\neq j$. The connections in the case of $n = 2$ are depicted in Figure \ref{mod_GenModel}. The variable $S_t$ describes the system that we analyze in the integrated information sense, hence this variable is taking the role of $X_t$ in the previous section. In the case of $\Phi_T$ the external influences are given by the actions $a_t$. 

\begin{center}
\begin{tikzpicture}
\draw[->, line width = 0.3mm] (0,0)--(1.5,0);

\draw[->, line width = 0.3mm] (2,0)--(3.5,0);

\draw[fill = white] (0.75,1.5) circle (0.365cm);
\draw[fill = white] (2.75,1.5) circle (0.365cm);

\draw[->, line width = 0.3mm] (0.5,0.65)--(2,0.65);
\draw[->, line width = 0.3mm] (2.5,0.65)--(4,0.65);

\draw[fill = white, white] (1.425,0.65) circle (0.05cm);
\draw[fill = white, white] (3.425,0.65) circle (0.05cm);

\draw[fill = white] (0,0) circle (0.365cm);
\draw[fill = white] (2,0) circle (0.365cm);
\draw[fill = white] (4,0) circle (0.365cm);

\draw[->, line width = 0.3mm] (0.5,0.65)--(0.35,-1);
\draw[->, line width = 0.3mm] (2.5,0.65)--(2.35,-1);
\draw[->, line width = 0.3mm] (4.5,0.65)--(4.35,-1);
\draw[->, line width = 0.3mm] (0,0)--(0.15,-1);
\draw[->, line width = 0.3mm] (2,0)--(2.15,-1);
\draw[->, line width = 0.3mm] (4,0)--(4.15,-1);

\draw[fill = white, white] (0.445,0.125) circle (0.05cm);
\draw[fill = white, white] (2.445,0.125) circle (0.05cm);

\draw[fill = white, white] (1.6,0.4) circle (0.05cm);
\draw[fill = white, white] (3.6,0.4) circle (0.05cm);

\draw[->, line width = 0.3mm] (0.75,1.5)--(2.1,0.75);
\draw[->, line width = 0.3mm] (2.75,1.5)--(4.1,0.75);

\draw[->, line width = 0.3mm] (0.75,1.5)--(1.7,0.3);
\draw[->, line width = 0.3mm] (2.75,1.5)--(3.7,0.3);

\draw[->, line width = 0.3mm] (0,0)--(1.5,0);
\draw[->, line width = 0.3mm] (2,0)--(3.5,0);

\draw[fill = white] (0.25,-1.5) circle (0.365cm);
\draw[fill = white] (2.25,-1.5) circle (0.365cm);
\draw[fill = white] (4.25,-1.5) circle (0.365cm);
\draw[fill = white] (0.75,1.5) circle (0.365cm);
\draw[fill = white] (2.75,1.5) circle (0.365cm);

\draw[fill = white] (0.5,0.65) circle (0.365cm);
\draw[fill = white] (2.5,0.65) circle (0.365cm);
\draw[fill = white] (4.5,0.65) circle (0.365cm);
\draw[] (0.5,0.65) node{$s^2_{t}$};
\draw[] (2.5,0.65) node{$s^2_{t+1}$};
\draw[] (4.5,0.65) node{$s^2_{t+2}$};

\draw[fill = white] (0,0) circle (0.365cm);
\draw[fill = white] (2,0) circle (0.365cm);
\draw[fill = white] (4,0) circle (0.365cm);

\draw[] (0,0) node{$s^1_{t}$};
\draw[] (2,0) node{$s^1_{t+1}$};
\draw[] (4,0) node{$s^1_{t+2}$};
\draw[] (0.25,-1.5) node{$o_{t}$};
\draw[] (2.25,-1.5) node{$o_{t+1}$};
\draw[] (4.25,-1.5) node{$o_{t+2}$};

\draw[->, line width = 0.3mm] (0.5,0.65)--(2,0.65);
\draw[->, line width = 0.3mm] (2.5,0.65)--(4,0.65);
\draw[->, dotted, line width = 0.3mm] (4.5,0.65)--(7.5,0.65);

\draw[fill = white, white] (1.425,0.65) circle (0.05cm);
\draw[fill = white, white] (3.425,0.65) circle (0.05cm);

\draw[fill = white] (0,0) circle (0.365cm);
\draw[fill = white] (2,0) circle (0.365cm);
\draw[fill = white] (4,0) circle (0.365cm);

\draw[->, line width = 0.3mm] (0.5,0.65)--(0.35,-1);
\draw[->, line width = 0.3mm] (2.5,0.65)--(2.35,-1);
\draw[->, line width = 0.3mm] (4.5,0.65)--(4.35,-1);
\draw[->, line width = 0.3mm] (8,0.65)--(7.85,-1);
\draw[->, line width = 0.3mm] (0,0)--(0.15,-1);
\draw[->, line width = 0.3mm] (2,0)--(2.15,-1);
\draw[->, line width = 0.3mm] (4,0)--(4.15,-1);
\draw[->, line width = 0.3mm] (7.5,0)--(7.65,-1);

\draw[fill = white, white] (0.445,0.125) circle (0.05cm);
\draw[fill = white, white] (2.445,0.125) circle (0.05cm);

\draw[->,  darkgray, line width = 0.3mm] (0.5,0.65)--(1.6,0.1);
\draw[->, darkgray, line width = 0.3mm] (0,0)--(2.05,0.51);
\draw[->, darkgray, line width = 0.3mm] (2.5,0.65)--(3.6,0.1);
\draw[->,  darkgray, line width = 0.3mm] (2,0)--(4.05,0.51);
\draw[->, dotted, darkgray, line width = 0.3mm] (5,0.65)--(7.1,0.1);
\draw[->, dotted, darkgray, line width = 0.3mm] (4.5,0)--(7.55,0.51);

\draw[fill = white, white] (1.6,0.4) circle (0.05cm);
\draw[fill = white, white] (3.6,0.4) circle (0.05cm);
\draw[->, line width = 0.3mm] (0.75,1.5)--(2.1,0.75);
\draw[->, line width = 0.3mm] (2.75,1.5)--(4.1,0.75);
\draw[->,line width = 0.3mm] (6.25,1.5)--(7.6,0.75);

\draw[->, line width = 0.3mm] (0.75,1.5)--(1.7,0.3);
\draw[->, line width = 0.3mm] (2.75,1.5)--(3.7,0.3);
\draw[->, line width = 0.3mm] (6.25,1.5)--(7.2,0.3);

\draw[fill = white, white] (0.425,0) circle (0.05cm);
\draw[fill = white, white] (2.425,0) circle (0.05cm);
\draw[->, line width = 0.3mm] (0,0)--(1.5,0);
\draw[->, line width = 0.3mm] (2,0)--(3.5,0);
\draw[->, dotted, line width = 0.3mm] (4,0)--(7,0);

\draw[fill = white] (0.25,-1.5) circle (0.365cm);
\draw[fill = white] (2.25,-1.5) circle (0.365cm);
\draw[fill = white] (4.25,-1.5) circle (0.365cm);
\draw[fill = white] (7.75,-1.5) circle (0.365cm);
\draw[fill = white] (0.75,1.5) circle (0.365cm);
\draw[fill = white] (2.75,1.5) circle (0.365cm);
\draw[fill = white] (6.25,1.5) circle (0.385cm);

\draw[fill = white] (0.5,0.65) circle (0.365cm);
\draw[fill = white] (2.5,0.65) circle (0.365cm);
\draw[fill = white] (4.5,0.65) circle (0.365cm);
\draw[fill = white] (8,0.65) circle (0.365cm);
\draw[] (0.5,0.65) node{$S^2_{1}$};
\draw[] (2.5,0.65) node{$S^2_{2}$};
\draw[] (4.5,0.65) node{$S^2_{3}$};
\draw[] (8,0.65) node{$S^2_{T}$};

\draw[fill = white] (0,0) circle (0.365cm);
\draw[fill = white] (2,0) circle (0.365cm);
\draw[fill = white] (4,0) circle (0.365cm);
\draw[fill = white] (7.5,0) circle (0.365cm);

\draw[] (0,0) node{$S^1_{1}$};
\draw[] (2,0) node{$S^1_{2}$};
\draw[] (4,0) node{$S^1_{3}$};
\draw[] (7.5,0) node{$S^1_{T}$};
\draw[] (0.25,-1.5) node{$O_{1}$};
\draw[] (2.25,-1.5) node{$O_{2}$};
\draw[] (4.25,-1.5) node{$O_{3}$};
\draw[] (7.75,-1.5) node{$O_{T}$};

\draw[] (0.75,1.5) node{$a_{1}$};
\draw[] (2.75,1.5) node{$a_{2}$};
\draw[] (6.25,1.5) node{$a_{\scriptscriptstyle T-1}$};
\end{tikzpicture}
    
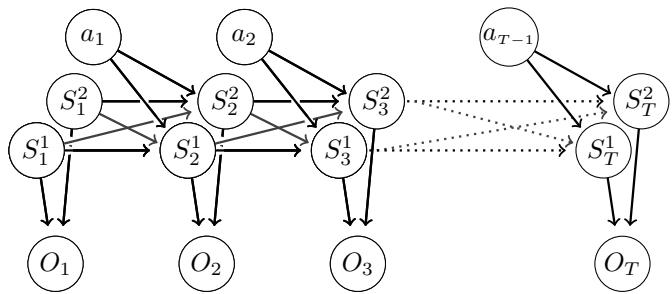
\captionof{figure}{Modified internal generative model in case of $n = 2$.} \label{mod_GenModel}
\end{center}

Our structural assumption is that the variables $S_t^i, S_t^j$ are conditionally independent at time $t$ given the full previous state $S_{t-1}$, with $i,j \in \{1,\ldots,n\}, i\neq j$, that is,
\begin{align}
    p(s_t \mid s_{t-1}, a_{t-1})
    = \prod_i p(s_t^i \mid s_{t-1}, a_{t-1}).
\end{align}
Thus, each component $S_t^i$ may depend on the entire previous state $S_{t-1}$.

Splitting the state space up into distinct parts is common in the active inference literature, where these parts are called state factors (See Appendix A in \cite{heins2022pymdp}). Often an even more restrictive assumption is made that different state factors are completely independent such that the forward model is given by 
\begin{align}
    p(s_t \mid s_{t-1}, a_{t-1}) = \prod_i p(s^i_t \mid s^i_{t-1}, a_{t-1}).
\end{align}
We however do not impose this stronger factorization across time, since cross-component dependencies via $s_{t-1}$ are essential for capturing integration effects. In Appendix \ref{app:act-inf} we discuss the modifications to the learning rules to enforce the conditional independence structure stated above.


\section{Experiment}\label{sect:setting}

We calculate the integrated information values of the simulated agents that learn to avoid touching the walls of their environment. The agents are trained via the active inference approach, as described in Section \ref{sect:activeinference} and defined in Appendix \ref{app:act-inf}. In the next section, we first describe the agents and their environment, and then define additional measures that we consider in the discussion of the results.

\subsection{Setting of the Experiment} 

The agents in our experiments should learn how to navigate a racetrack environment while avoiding to touch the walls.
The Figure \ref{Fig:Experiment} on the left depicts a sketch of such an agent. It consists of a round body with a tail that marks the back and two binary sensors that can detect the walls of the environment. The agent can move as if it had two wheels that can spin either fast or slow, so it can go fast forward, slow forward, forward to the left or forward to the right.

\begin{center}
    \includegraphics[height = 0.25\textheight]{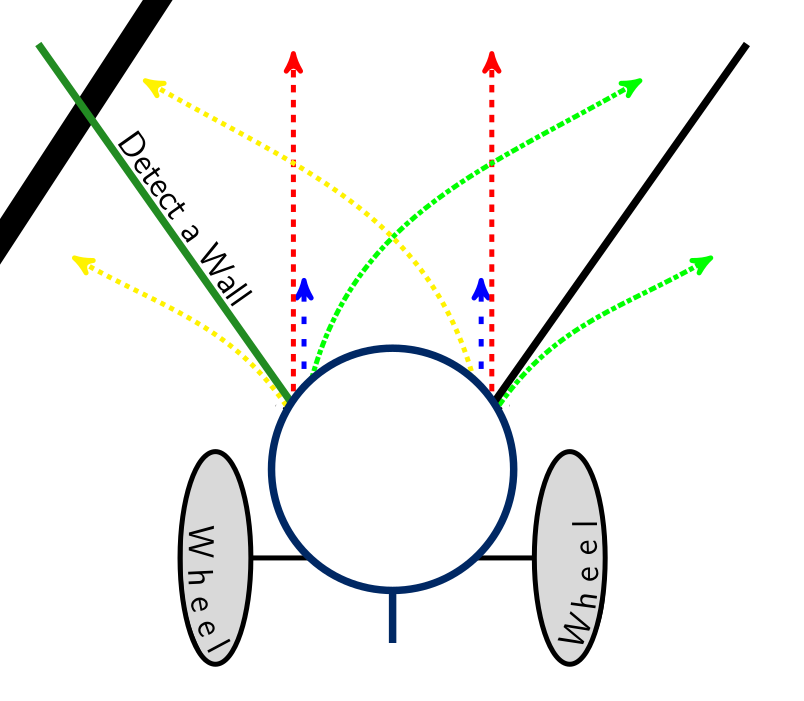} \hfill
        \includegraphics[height = 0.25\textheight]{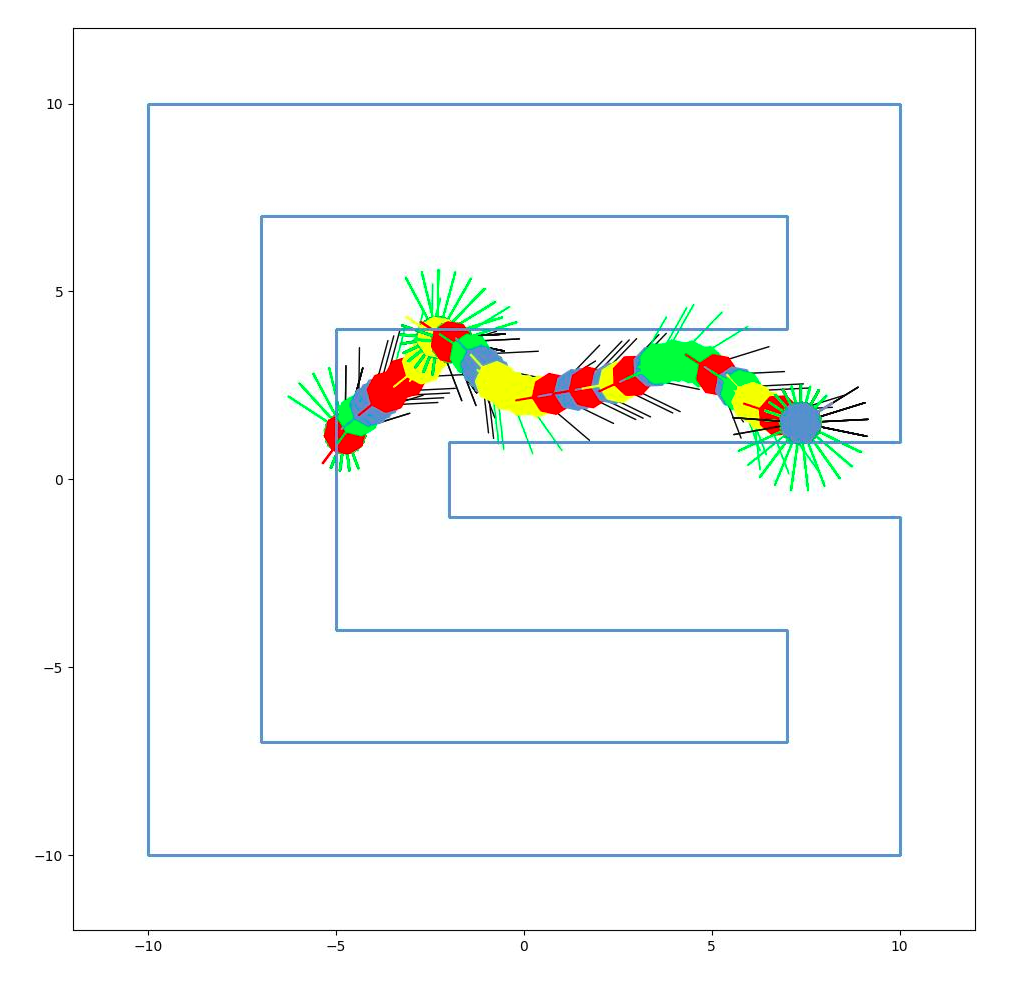} \hspace{1cm}
    \captionof{figure}{Sketch of an agent detecting a wall on the left, a trajectory of an agent in its environment on the right. } \label{Fig:Experiment}
\end{center}

The right hand side of Figure \ref{Fig:Experiment} shows how an agent moves in its environment. A video of an agent moving in this environment can be found at \cite{Langer_git_2026}. The agent moves multiple steps through the environment and in each step the color of the body indicates what action is chosen for the next movement. If the body of the agent touches a wall, then this agent is stuck and can only turn on the spot. Once both sensors do not detect a wall anymore, so once the agent faces a way from the wall, the agent is free to move away from it. 

The goal of not touching the wall is included in the action selection algorithm by setting the preference distribution in the expected free energy in equation \eqref{eq:efe} to favor observations in which the body of the agent does not touch a wall.

The changes to the learning algorithm, described in Section \ref{sect:combining}, would in theory be sufficient to train an active inference agent on which the calculation of integrated information measures is possible. However, we observe that performing the updates of all three sets of parameters as described in \eqref{eq:approx-update-D}, \eqref{eq:approx-update-A} and \eqref{eq:approx-update-B} at the same time does in general not lead to agents with a representation of their environment. In that case, the agents learn to repeat one turn movement, either left or right, without any variation, hence the observations $O_t$ and the internal states $S_t$ become independent. Since these agents are not successful in the sense of the active inference framework, we reduce the complexity of the task by sampling the distribution $p(O_t \vert S_t)$. We discuss alternative solutions to this problem in Section \ref{sect:discussion}.

\subsection{Success Rate and Multiinformation}

In addition to the free energy and the various integrated information measures, we also consider two additional values: the success rate and the multiinformation of the generative model. 

The success rate is a sampled quantity, where we simply save and then normalize how often an agent's body is touching a wall. Hence, a success rate of 0.1 translates to an agent that was stuck 90$\%$ of the 50\,000 total steps. An agent with a high success rate has therefore learned to effectively avoid the walls. 

The multiinformation is an information-geometric quantity that is also known as total correlation and can be seen as a generalization of the mutual information. Its properties and the relationship to stochastic dependence are discussed in \cite{Studený1998}.

Let $X$ be a multivariate, discrete random variable with $X = (X_1, \dots, X_n)$, then the multiinformation is defined as follows:
    \begin{equation*}
                MI(X) = \sum\limits_{i = 1}^n H(X_i) - H(X_1, \dots , X_n )
    \end{equation*}
    where $H$ denotes the entropy $H(X_i) = - \sum\limits_{x_i} p(x_i) \log p(x_i)$.

In the experiments we calculate the multiinformaion of the generative model, hence \\ $X = (O_{t+1}, S_{t+1}, A_{t+1}, S_t, A_t)$, in order to assess the total correlation of the information flow inside the generative model. 

\section{Results} \label{sect:results}

Here, we discuss the results of the experiments. Each agent performs 50\,000 steps in the environment and we save a data point consisting of the different measure every 1\,000 steps. Additionally, we calculate the VFE, as described in \ref{VFE}, every 1000 steps for the duration of 100 steps and then average over these 100 steps. This is then taken as an approximation of the free energy. We discuss this relationship between this free energy and other potential measures for the prediction error in Appendix \ref{App:B}. 
The agents that we analyze differ in the number of internal states, more precisely, we train agents with $n =2, 3$ and $4$ binary internal nodes. 

For each type of agent we perform 1000 such tests and calculate the Pearson and Spearman correlations between them using the Fisher z-transformation. There we compute the Pearson or Spearman correlation among the different measures for each time step, perform a z-transformation and calculate the arithmetic mean over the resulting z-values. Finally, we transform the values back. This method is described in, for example, \cite{fisher1915frequency,silver1987averaging}.
Since the agents first need to learn and build their generative model, we analyze the data starting at step 25000. However, the full results are depicted in Appendix \ref{ap:results}. 

\begin{center}
    \includegraphics[width = 0.485\textwidth]{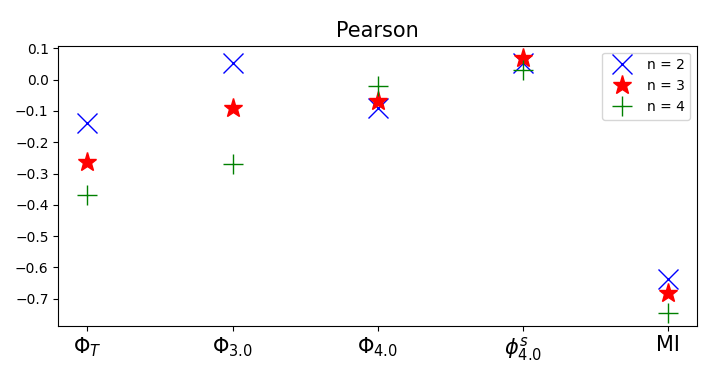}
        \includegraphics[width = 0.485\textwidth]{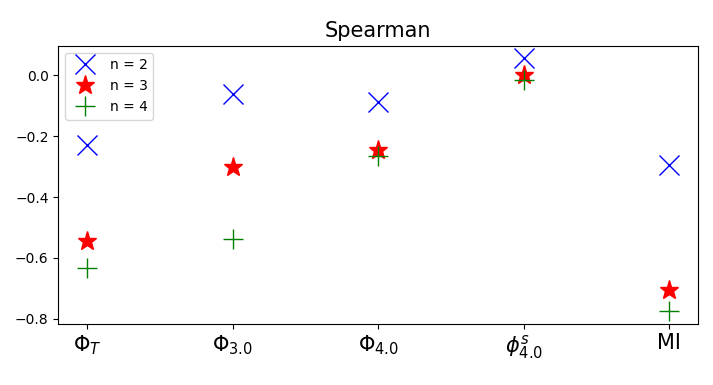}
    \captionof{figure}{Spearman correlation results between the free energy and the integrated information measures.} \label{fig:results1}
\end{center}

In Figure \ref{fig:results1} we see the results of the different integrated information measures and the multiinformation (MI) correlated to the free energy, with the Pearson correlation on the left and the Spearman correlation on the right.
The correlations between the free energy and the integrated information measures, as well as the MI, strengthen with the size of the hidden states. We observe that $\Phi_T$ and the MI are highly negatively correlated to the free energy for $n = 4$ using the Spearman correlation. Hence, the lower the free energy, the higher $\Phi_T$ and the MI.

We can observe a similar trend for $\Phi_{3.0}$ with an increase of the number of nodes, but the correlations are overall slightly weaker. Since the calculations for IIT 3.0 and 4.0 are very costly, it is unfortunately not possible to increase the size of $n$ much further. There are many differences in the calculations of the various measures, as briefly discussed in Section \ref{sect:IIT}, most notably the dependencies of $\Phi_{3.0}$ and $\Phi_{4.0}$ on the current state and their focus on only the internal nodes. Hence, we have additionally calculated the Spearman correlation between these measures and the negative log evidence $- \log p(o_t)$, which is the state-dependent value that we average to get to the free energy. This calculation also shows only a very weak negative correlation between the negative log likelihood and $\Phi_{3.0}$, as well as $\Phi_{4.0}$. These values lie roughly between 0 and -0.2 in the case of $\Phi_{4.0}$, around 0 for $\phi^s_{4.0}$ and between 0 and -0.3 for $\Phi_{3.0}$.

Now we take a closer look at the correlations among the IIT measures. Figure \ref{fig:IITmeasures} depicts in the first row the Pearson correlation matrices among the integrated information related measures for $n = 2, 3$ and $4$ and in the second row the Spearmon correlation matrices. 
We observe that $\Phi_{3.0}$ and $\phi_{4.0}^s$ are correlated to $\Phi_{4.0}$  and that the correlation to $\Phi_T$ and the MI is weaker. In Section \ref{ap:results} we depict the complete results of the experiments and there we observe that the different measures result in vastly different scales.  
Considering the many differences in the calculations of these measures, even a weak correlation is surprising and an argument for some level of consistency in the theory that extends from the axioms to the mathematical formulation.

\begin{center}  
\includegraphics[width = 0.325\textwidth]{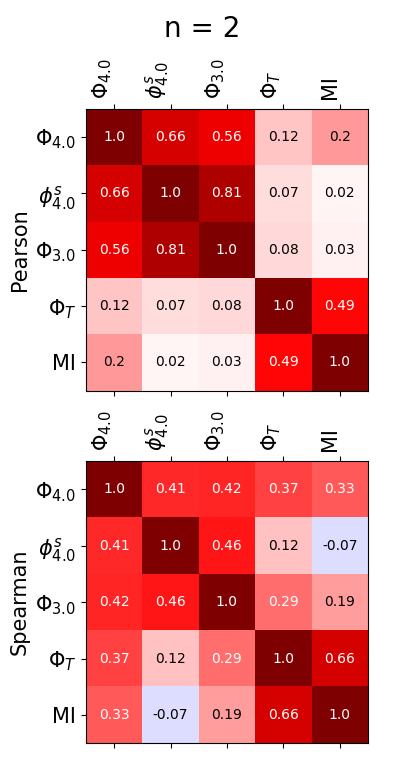} 
\includegraphics[width = 0.325\textwidth]{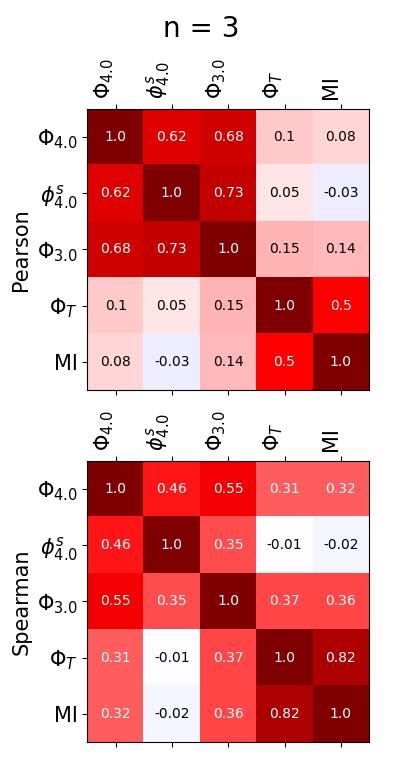}
\includegraphics[width = 0.325\textwidth]{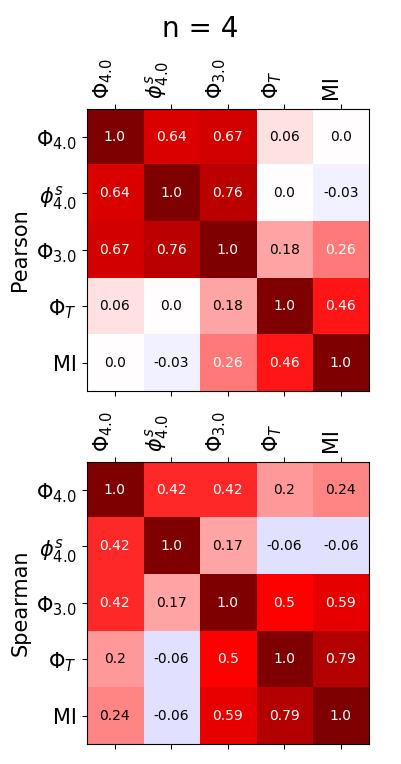}
\captionof{figure}{Pearson and Spearman correlation among the IIT related measures.} \label{fig:IITmeasures}
\end{center}

It is important to note that agents that are successful at minimizing free energy do not necessarily perform well in terms of the success rate. The correlation between the success rate and the free energy results in approximately 0.01, -0.08 and -0.11 for Pearson and 0.02, -0.06 and -0.07 for Spearman and $n=2,3$ and $4$ respectively. In Appendix \ref{ap:results} we observe that there is a small number of agents that have a very low free energy, but a success rate close to 0. Removing agents with a final success rate below 2$\%$ leads to correlation coefficients of roughly -0.23, -0.31, -0.2 for Pearson and -0.1, -0.33 and -0.27 for Spearman. Hence, being good at minimizing the free energy does not necessarily lead to the desired behavior.

\section{Discussion} \label{sect:discussion}

In this article we combine the active inference framework and the Integrated Information theory and examine their relationship empirically. In our experiments we observe that the measures $\Phi_T$ and the MI are much closer to assessing the information flow in the network compared to the integrated information measures from IIT 3.0 and 4.0. The latter two have been derived considering further axioms in addition to \enquote{information} and \enquote{integration}. Hence, there seems to be a relation between the information integration in a system and the minimization of the free energy, but not necessarily a relation between the amount of consciousness, as defined by IIT, and the free energy.

While training the agents we encountered the problem that learning all the different parts of the generative model at once leads to a convergence to an undesirable local optimum. There the observation is independent of the internal states. In our experiments we decided to make the task easier by sampling the distribution $p(O_t,S_t)$ and thereby reducing the number of parameters that need to be optimized via active inference.  
An alternative approach to learning the whole generative model is proposed by the authors of \cite{DaCosta}. They suggest that the different aspects of the active inference framework follow different time scales. Hence, we could vary the number of time steps after which we update the parameters defined in \eqref{eq:approx-update-D}--\eqref{eq:approx-update-B}.
 
Additionally, more involved agents or tasks might reveal additional nuances in the relationship between these theories. However, the measures defined in IIT 3.0 and 4.0 are computationally expensive, which limits the size of the generative model.

Furthermore, a more detailed theoretical treatment of the relationship among the different measures could advance the understanding of their relationship. Here a joint theoretical treatment of the measures $\Phi_T$, MI and the free energy seems to be natural, whereas the mathematically vastly different definitions of the measures in IIT 3.0 and 4.0 make a theoretical comparison more difficult. 

\section*{Acknowledgments}
The authors would like to thank Lancelot Da Costa, Pablo Lanillos, Thomas Parr and Larissa Albantakis for the helpful discussions and William Mayner for help with the PyPhi toolbox.

We acknowledge the support of the Deutsche Forschungsgemeinschaft project 402780474 (Information Integration in Predictive Processes: A Mechanistic Grounding of the Self) within the Priority Programme “The Active Self” (SPP 2134). 

\printbibliography
\appendix

\section{Active Inference} \label{app:act-inf}

\subsubsection*{Action selection}
Let an agent be at time step $t$, having received observations $o_{1:t}$ and performed actions $a_{1:t-1}$. The agent selects its next action by sampling a policy $\pi_t$, which is a sequence of future actions $(a_t, \cdots a_T)$, from the following distribution: 
\begin{align} \label{eq:approx-post-policies-1}
   \sigma(- G(\pi_t| o_{1:t}, a_{1:t-1}) )
\end{align}
and selecting the action $a_t$ corresponding to that policy. The function $\sigma$ denotes the softmax function and $G$ is the expected free energy function given by
\begin{align}   
    \begin{aligned} 
        G(\pi_t| o_{1:t}, a_{1:t-1}) = - \bigg(&\bE_{q_t(O_{t+1:T}| \pi_t)}\bigg[\DKL \Big( q_t(S_{t+1:T} | o_{t+1:T}, \pi_t) \parallel q_t(S_{t+1:T} |\pi_t) \Big) \bigg] \\ 
        &+ \bE_{q_t(O_{t+1:T} | \pi_t)} \Big[\ln p_C(O_{t+1:T})\Big]\bigg), \label{eq:efe}
    \end{aligned}
\end{align}
where $p_C$ is a preference distribution over observations that is given explicitly to the agent. 

\subsubsection*{Perception}
For a general $t$ the belief about $s_t$ is updated as follows:
\begin{align} \label{eq:update-state}
    q(S_t| o_{1:t},  a_{1:t-1}) \propto p(O_t|s_t) \sum_{s_{t-1}} p(S_t|s_{t-1}, a_{t-1}) q(s_{t-1}|o_{1:t-1}, a_{1:t-2}).  
\end{align}
For future time point $\tau > t$, the belief about the state  $s_\tau$ is given by
\begin{align} \label{eq:posterior-future-state}
    q(S_\tau| o_{1:t}, a_{1:\tau-1} ) = \sum_{s_{\tau-1}} p(S_\tau| s_{\tau-1}, a_{\tau-1})  q(s_{\tau-1}| o_{1:t}, a_{1:\tau-2}).
\end{align}
The belief about a future state given a future observation is calculated as follows:
\begin{align}  \label{eq:posterior-future-state-given-obs}
    q(S_\tau| o_\tau, o_{1:t},  a_{1:\tau-1} ) \propto p(o_\tau | S_\tau)  q(S_\tau| o_{1:t},  a_{1:\tau-1} ). 
\end{align}
In order to compute $G$ in \eqref{eq:efe}, we need a posterior distribution $q(O_\tau| o_{1:t}, a_{1:\tau-1})$ over future observations $o_\tau$. This can be computed as follows:
\begin{align} \label{eq:posterior-future-obs}
    q(O_\tau|o_{1:t}, a_{1:\tau-1}) = \sum_{s_\tau} p(O_\tau|s_\tau) q(s_\tau| o_{1:t},  a_{1:\tau-1} ).
\end{align}

In order to perform inference over states in the past, present and future (which is needed for the learning of the generative model and for the computation of the variational free energy over states), the agent can use the following formula:
\begin{align}
    q(S_{1:T}| o_{1:t}, a_{1:t-1}, \pi_t) &\propto p(
    o_{1:t}|S_{1:T}, a_{1:t-1}, \pi_t) p(S_{1:T} | a_{1:t-1}, \pi_t) \\
    &=  p(o_{1:t}|S_{1:t}) p(S_{1:T} | a_{1:t-1}, \pi_t) \\
    &= \prod_{\tau=1}^t p(o_\tau | S_\tau) \ p(S_1) \prod_{\tau=2}^T p(S_\tau|S_{\tau-1}, a_{\tau-1}), \label{eq:posterior-states-past-future}
\end{align}
where we use $a_{\tau-1}$ in the last term for elements of both $a_{1:t-1}$  and $\pi_t$. 

\subsubsection*{Learning}

The generative model consists of three (conditional) categorical distributions that are parametrized by $\theta = (\theta^D, \theta^A, \theta^B)$ in the following way:
\begin{align}
    &p(s_1^{(j)}| \theta^D) = \theta^D_j, \label{gen-mod1} \\
    &p(o_\tau^{(i)}| s_\tau^{(j)}, \theta^A) = \theta^A_{ij}, \label{gen-mod2} \\
    &p(s_{\tau}^{f (j)}| s_{\tau-1}^{(k)}, a_{\tau-1}^{(l)}, \theta^B) = \theta^{B}_{fjkl}, \label{gen-mod3}
\end{align}
where we have enumerated the elements of the observation, action and state space with the bracketed superscript $^{(\cdot)}$, and we use normal superscript $^f$ to index the different state factors / dimensions. 
In order to learn the parameters of the generative model, we adopt a Bayesian belief updating scheme with a Dirichlet prior.  More specifically, the prior over $\theta$ is parametrized by $\alpha = (\alpha^D, \alpha^A, \alpha^B)$ and is given by 
\begin{align}
    p(\theta|\alpha) &= p(\theta^D | \alpha^D) \prod_j p(\theta^A_{\bullet j} | \alpha^A) \prod_{f,k,l} p(\theta^B_{f \bullet kl} | \alpha^B) \label{gen-mod-prior1}\\
    p(\theta^D | \alpha^D) &\propto  \prod_j \left(\theta^D_j\right)^{\alpha^D_j - 1}, \label{gen-mod-prior2} \\
    p(\theta^A_{\bullet j} | \alpha^A) &\propto  \prod_i \left(\theta^A_{ij}\right)^{\alpha^A_{i j} - 1}, \label{gen-mod-prior3}\\
    p(\theta^B_{f \bullet kl} | \alpha^B) &\propto  \prod_j \left(\theta^B_{fjkl}\right)^{\alpha^B_{fjkl} - 1}, \label{gen-mod-prior4}
\end{align}
where $\theta_{\bullet j}$ denotes the vector $(\theta_{1j},\ldots, \theta_{nj})$.

Now after performing actions $a_{1:T-1}$ and receiving observations $o_{1:T}$ we want to update our belief about $\theta$ according to Bayes' rule. The approximate posterior distribution is a Dirichlet distribution where the hyperparameter $\alpha$ is updated in the following way:
\begin{align}
    {\alpha^D_j}'  &= \alpha^D_j + q_T\left(s_1^{(j)}\right), \label{eq:approx-update-D}\\
    {\alpha^A_{ij}}'  &= \alpha^A_{ij} + \sum_{\tau=1}^T \mathbbm{1}_{o^{(i)}}(o_\tau) q_T\left(s_\tau^{(j)}\right), \label{eq:approx-update-A}\\
    {\alpha^B_{fjkl}}'  &= \alpha^B_{fjkl} + \sum_{\tau=2}^{T} q_T\left(s_{\tau}^{f(j)}\right) q_T\left(s_{\tau-1}^{(k)} \right)  \mathbbm{1}_{a^{(l)}}(a_{\tau-1}). \label{eq:approx-update-B}
\end{align}
The distributions $q_T\left(s_\tau\right), \tau \in \{1,\ldots,T\}$ are approximate posteriors obtained using the current version of the generative model (before $\theta$ has been updated).  

In  order to go from a Dirichlet distribution $p(\theta | \alpha)$ to an actual value of the parameter that can be used for the generative model, the mean of the distribution is used, which is given by
\begin{align}
    \hat{\theta}_i &= \bE_{p(\theta | \alpha)}[\theta_i] \\
    &= \frac{\alpha_i}{\sum_j \alpha_j}. \label{eq:update-theta}
\end{align}
\noeqref{eq:update-theta}

\newpage
\section{Relationship between the free energy and other measures for the prediction error} \label{App:B}

The VFE, as defined in \eqref{VFE}, is a state-dependent quantity that has to be calculated for each step of the agent and changes not only with the observations, but also with the belief $q(O_t \vert o_{1:t-1}, a_{1:t-1})$. Hence, for more advanced applications it would be useful to have a measure that is easier to calculate and approximates the behavior of the variational free energy.

In the Figure \ref{fig:cand_VFE}, we depict the Pearson correlation of the different candidates to the free energy, as an average over the VFE, on the left and the Spearman correlation on the right. Overall we observe that the correlations, positive or negative, increase with the number of the internal nodes.

The first measure is the KL-divergence between the sampled observations $\tilde{p}(O_t)$ and the internal belief over the observations $p(O_t) = \sum\limits_{s_t} p(O_t \vert s_t) q(s_t)$, where $q(S_t)$ is the belief over the current state. Here we denote by $O^{\star}$ the random variable corresponding to a sampled distribution $\tilde{p}$. Then we have 
\begin{align*}
    D(O^{\star}_t \parallel O_t) = \sum\limits_{o_t} \tilde{p}(o_t) \log \dfrac{\tilde{p}(o_t)}{ p(o_t)} = H_{\tilde{P}, P}(O_t) - H(O^{\star}_t),
\end{align*}
where $H_{\tilde{P}, P}(O_t)$ is the cross-entropy 
\begin{align*}
    H_{\tilde{P}, P}(O_t) = - \sum\limits_{o_t} \tilde{p}(o_t) \log p(o_t).
\end{align*}
In Figure \ref{fig:cand_VFE}, we observe that $D(O^{\star}_t \parallel O_t)$, as well as $H_{\tilde{P}, P}(O_t)$ (depicted as the last measure), are strongly inversely correlated to the free energy, while $D( O_t \parallel O^{\star}_t)$ is slightly positively correlated and $H_{ P, \tilde{P}}(O_t)$ is positively correlated to the free energy. Hence, the measures where the internal prediction $p$ 
is in front of the logarithm lead to a positive correlation, while $\tilde{p}$ leads to a negative correlation.

\begin{center}
      \includegraphics[width = 0.475\textwidth]{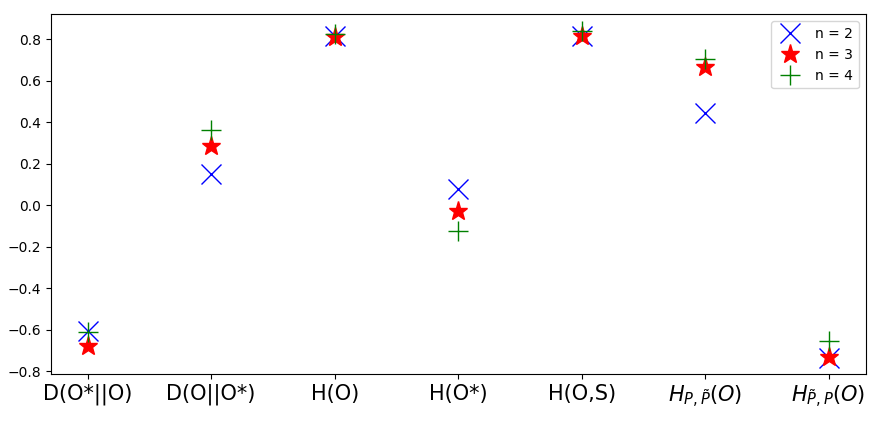} 
            \includegraphics[width = 0.475\textwidth]{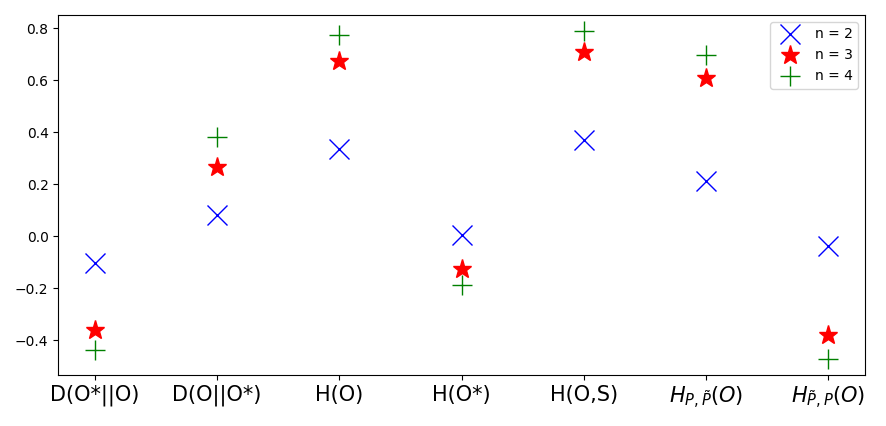}
      \captionof{figure}{ Pearson correlation to the free energy on the left an Spearman on the right.} \label{fig:cand_VFE}
\end{center}

The VFE is minimized if the model learns to internally assign a high probability to the observations that actually occur $q(o_t^{\star} \vert a_{t-1}, o_{t-1})$. Hence, it is not aimed at learning the overall distribution of the observations, but to accurately predict the next observation. In each time step the best distribution $q$ would be the one that assigns a likelihood of 1 to the observation that occurs. The KL-divergences and cross entropies measure, however, how close overall the distributions are to each other, which is a slightly differnt objective.

The entropies $H(O_t,S_t)$ and $H(O)$ that purely consider the internal generative model, without any direct influence of the actual sampled observations, lead to the strongest correlation with the free energy. The lower the entropy, the more structure the distribution has and the better the VFE. 

        \newpage
\section{Complete Results} \label{ap:results}
Here we depict the data from the experiments. In the following section the results for $n =2$ is depicted in the first, $n =3$ in the second and $n =4$ in the last column. In each figure the x-axis signifies the steps that the agents take and on the y-axis are the 1000 agents depicted. The first row of results depicts the free energy. The agents are sorted via the arithmetic mean of their free energy values. Hence, the agent at position 0 has overall the lowest free energy value, while the agent as position 999 has the highest. In all following figures the agents remain to be sorted in the same way, so the agent with the lowest overall free energy is depicted at position 0 in all figures. 

\begin{table}[h]
    \centering
    \begin{tabular}{p{0.05cm} p{0.05cm} c c c}
     &  &  $n = 2$  & $n = 3$ & $n = 4$  \\
     \raisebox{7\normalbaselineskip}[0pt][0pt]{\rotatebox[origin=c]{90}{Free Energy}} &      \raisebox{7\normalbaselineskip}[0pt][0pt]{\rotatebox[origin=c]{90}{\small agents}} &    \includegraphics[width = 4.25cm]{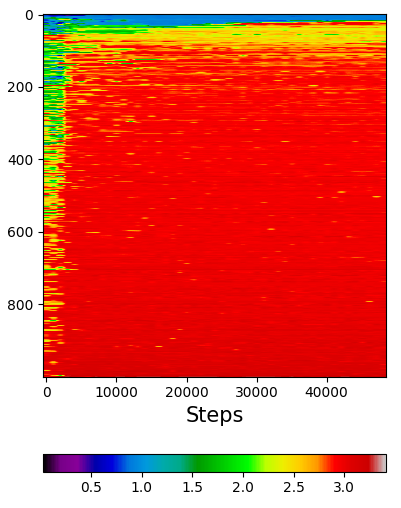}   &         \includegraphics[width = 4.25cm]{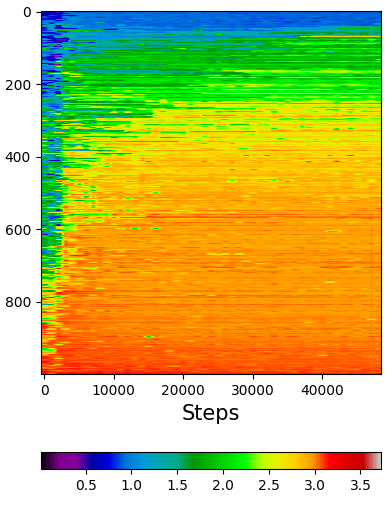} &  
            \includegraphics[width = 4.25cm]{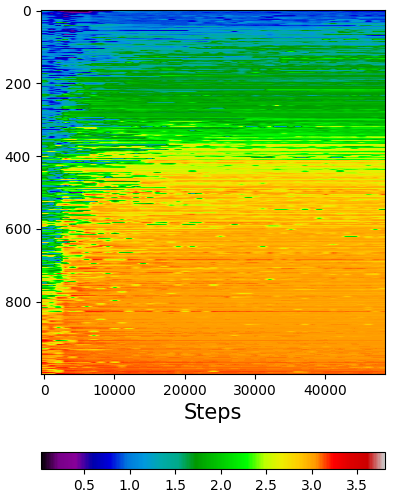} \\
     \raisebox{7\normalbaselineskip}[0pt][0pt]{\rotatebox[origin=c]{90}{Success Rate}} &      \raisebox{7\normalbaselineskip}[0pt][0pt]{\rotatebox[origin=c]{90}{\small agents}} &  
           \includegraphics[width = 4.25cm]{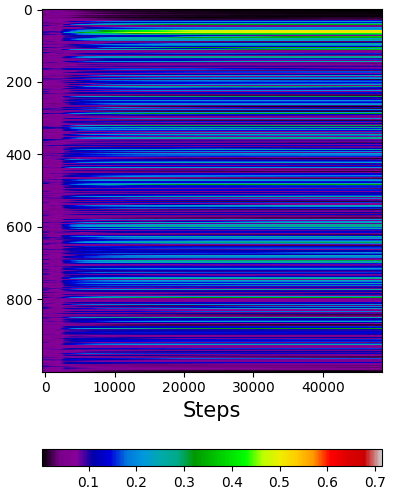}   &         \includegraphics[width = 4.25cm]{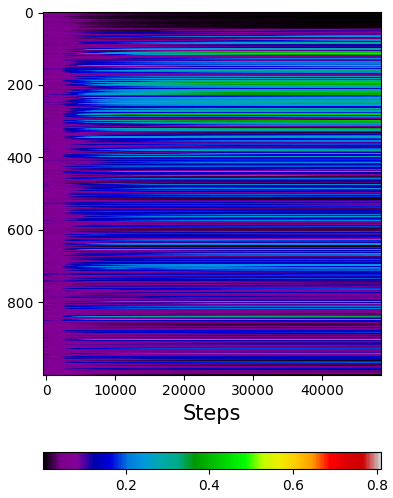} &  
            \includegraphics[width = 4.25cm]{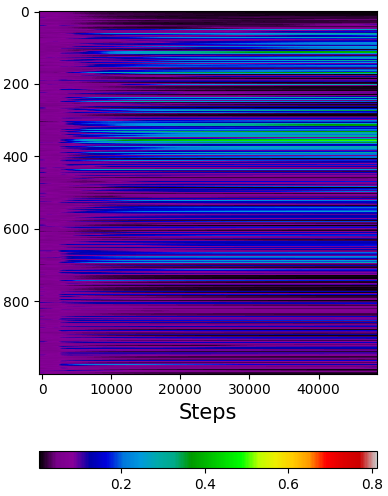}   \\
     \raisebox{7\normalbaselineskip}[0pt][0pt]{\rotatebox[origin=c]{90}{Multiinformation}} &      \raisebox{7\normalbaselineskip}[0pt][0pt]{\rotatebox[origin=c]{90}{\small agents}} &  
           \includegraphics[width = 4.25cm]{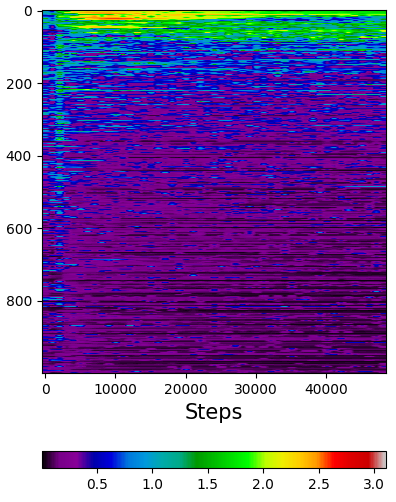}   &         \includegraphics[width = 4.25cm]{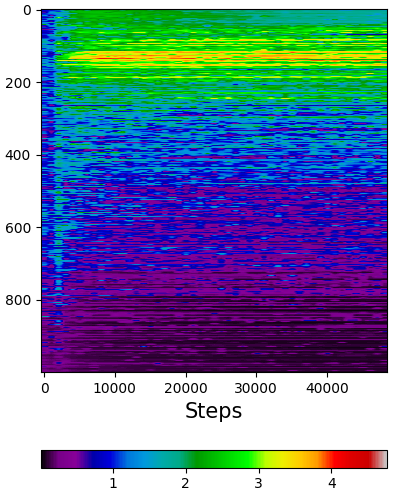} &  
            \includegraphics[width = 4.25cm]{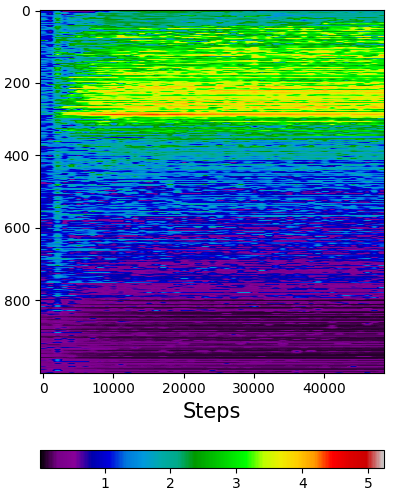}  
    \end{tabular}
\end{table}

\begin{table}[h]
    \centering
    \begin{tabular}{p{0.05cm} p{0.05cm} c c c}
     &  &  $n = 2$  & $n = 3$ & $n = 4$  \\
     \raisebox{7\normalbaselineskip}[0pt][0pt]{\rotatebox[origin=c]{90}{$\Phi_T$}} &      \raisebox{7\normalbaselineskip}[0pt][0pt]{\rotatebox[origin=c]{90}{\small agents}} &    \includegraphics[width = 4.25cm]{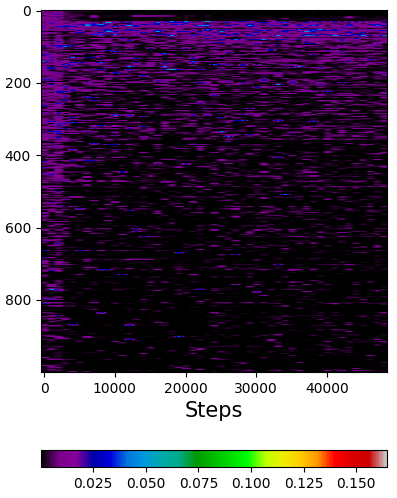}   &         \includegraphics[width = 4.25cm]{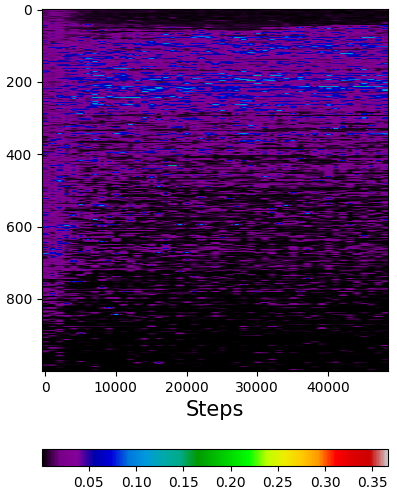} &  
            \includegraphics[width = 4.25cm]{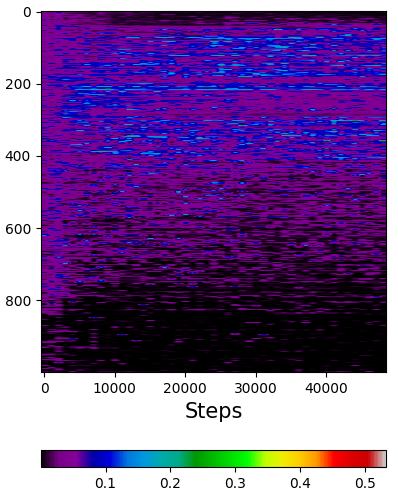} \\
     \raisebox{7\normalbaselineskip}[0pt][0pt]{\rotatebox[origin=c]{90}{$\Phi_{3.0}$}} &      \raisebox{7\normalbaselineskip}[0pt][0pt]{\rotatebox[origin=c]{90}{\small agents}} &  
           \includegraphics[width = 4.25cm]{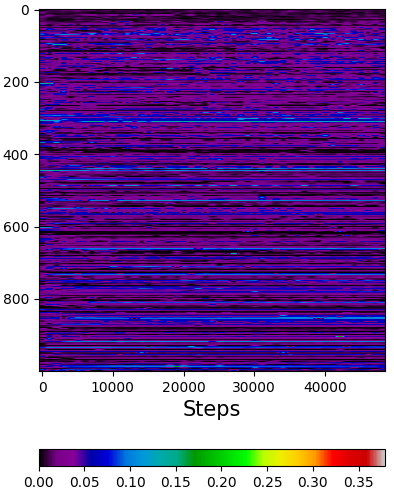}   &         \includegraphics[width = 4.25cm]{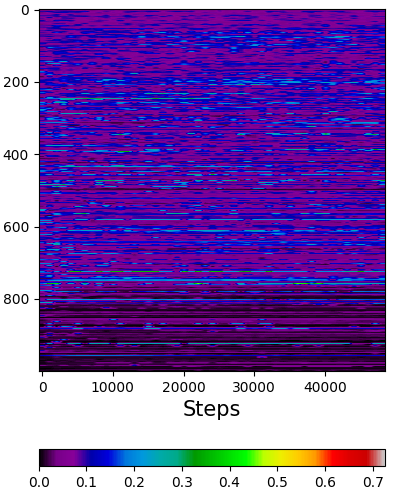} &  
            \includegraphics[width = 4.25cm]{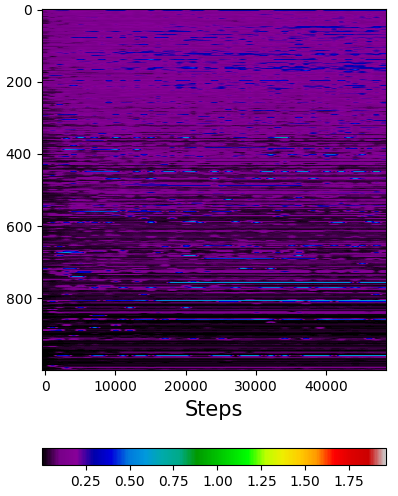}   \\
     \raisebox{7\normalbaselineskip}[0pt][0pt]{\rotatebox[origin=c]{90}{$\Phi_{4.0}$}} &      \raisebox{7\normalbaselineskip}[0pt][0pt]{\rotatebox[origin=c]{90}{\small agents}} &  
           \includegraphics[width = 4.25cm]{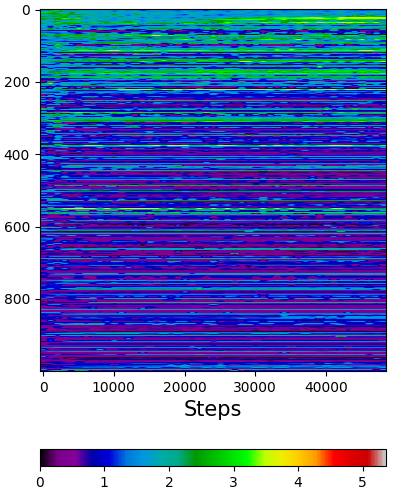}   &         \includegraphics[width = 4.25cm]{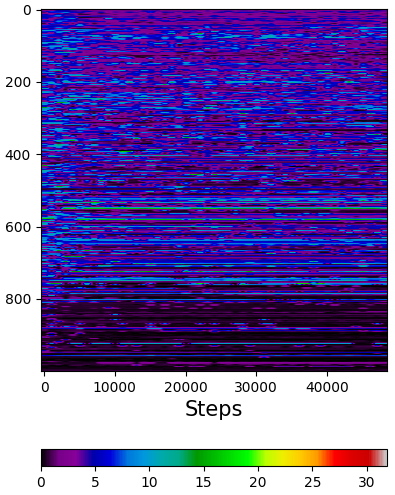} &  
            \includegraphics[width = 4.25cm]{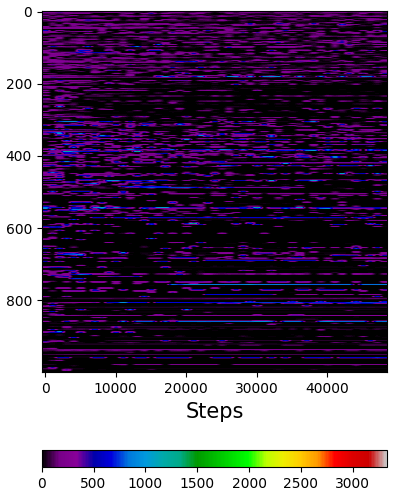}  
    \\
     \raisebox{7\normalbaselineskip}[0pt][0pt]{\rotatebox[origin=c]{90}{$\phi_{4.0}^s$}} &      \raisebox{7\normalbaselineskip}[0pt][0pt]{\rotatebox[origin=c]{90}{\small agents}} &  
           \includegraphics[width = 4.25cm]{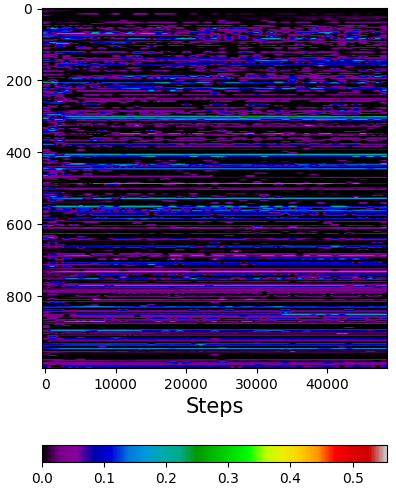}   &         \includegraphics[width = 4.25cm]{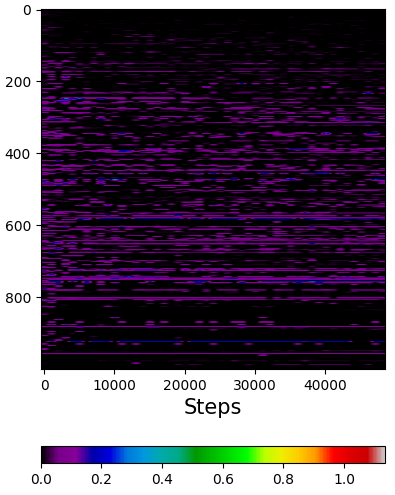} &  
            \includegraphics[width = 4.25cm]{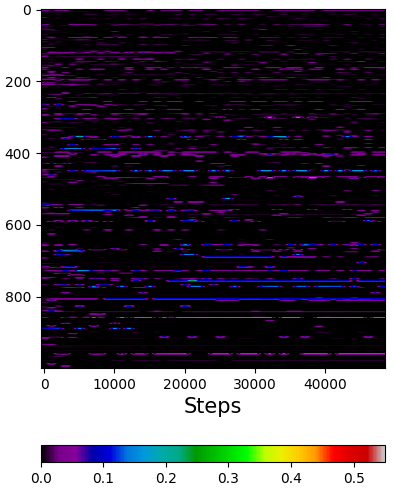}  
    \end{tabular}
\end{table}

\begin{table}[h]
    \centering
    \begin{tabular}{p{0.05cm} p{0.05cm} c c c}
     &  &  $n = 2$  & $n = 3$ & $n = 4$  \\
     \raisebox{7\normalbaselineskip}[0pt][0pt]{\rotatebox[origin=c]{90}{ $D(O^{\star} \parallel O)$}} &      \raisebox{7\normalbaselineskip}[0pt][0pt]{\rotatebox[origin=c]{90}{\small agents}} &    \includegraphics[width = 4.25cm]{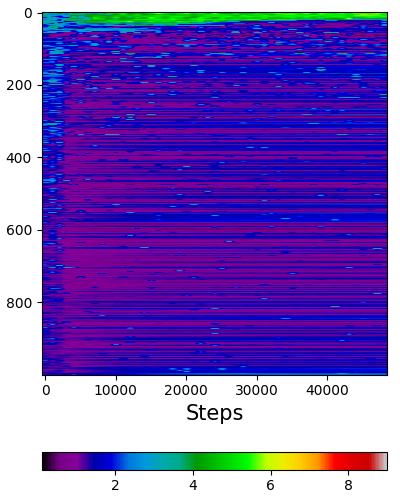}   &         \includegraphics[width = 4.25cm]{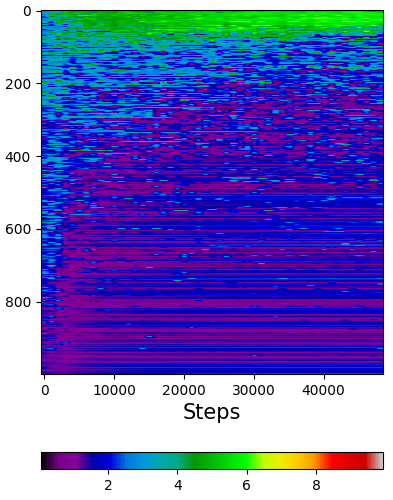} &  
            \includegraphics[width = 4.25cm]{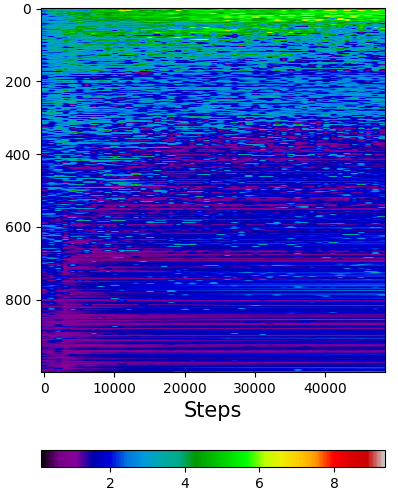}
                \\
     \raisebox{7\normalbaselineskip}[0pt][0pt]{\rotatebox[origin=c]{90}{ $D(O \parallel O^{\star})$}} &      \raisebox{7\normalbaselineskip}[0pt][0pt]{\rotatebox[origin=c]{90}{\small agents}} &  
           \includegraphics[width = 4.25cm]{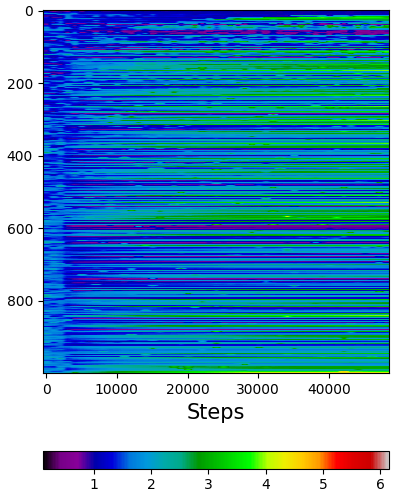}   &         \includegraphics[width = 4.25cm]{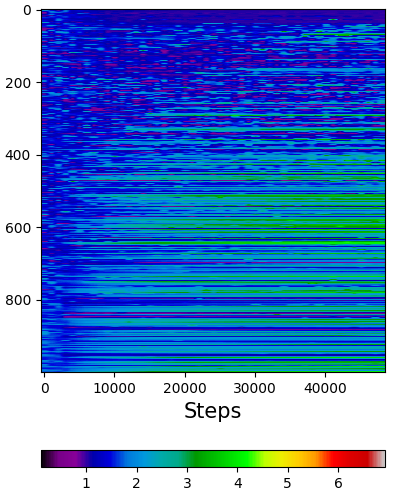} &  
            \includegraphics[width = 4.25cm]{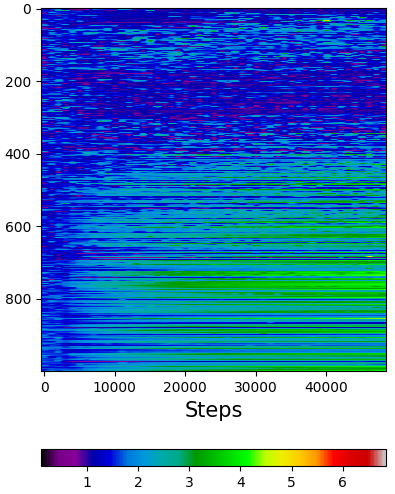}  \\
     \raisebox{7\normalbaselineskip}[0pt][0pt]{\rotatebox[origin=c]{90}{ $H_{ \tilde{P}, P}(O)$}} &      \raisebox{7\normalbaselineskip}[0pt][0pt]{\rotatebox[origin=c]{90}{\small agents}} &  
           \includegraphics[width = 4.25cm]{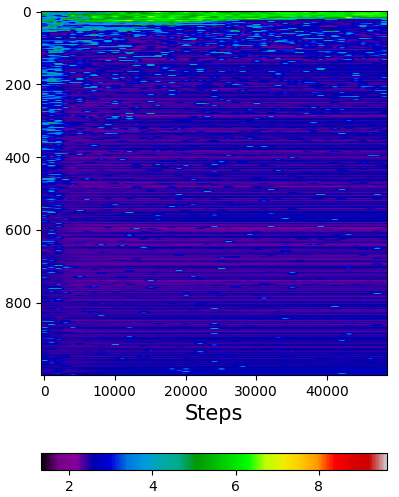}   &         \includegraphics[width = 4.25cm]{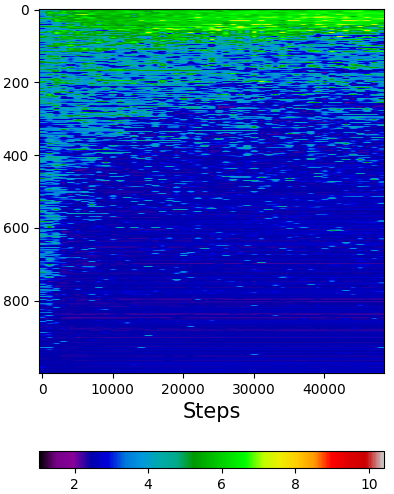} &  
            \includegraphics[width = 4.25cm]{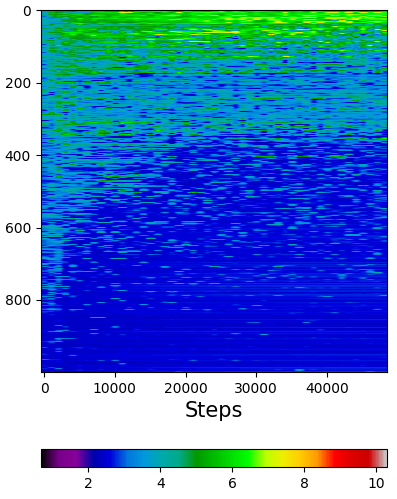}  \\
     \raisebox{7\normalbaselineskip}[0pt][0pt]{\rotatebox[origin=c]{90}{$H_{P, \tilde{P}}(O)$}} &      \raisebox{7\normalbaselineskip}[0pt][0pt]{\rotatebox[origin=c]{90}{\small agents}} &  
           \includegraphics[width = 4.25cm]{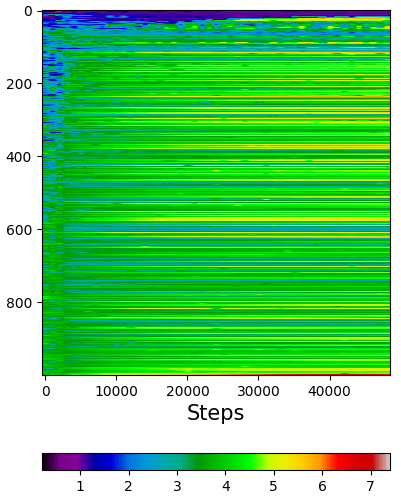}   &         \includegraphics[width = 4.25cm]{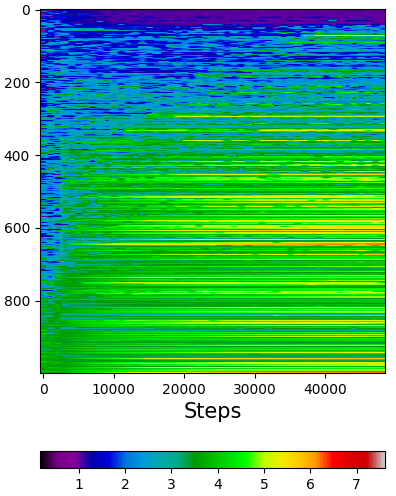} &  
            \includegraphics[width = 4.25cm]{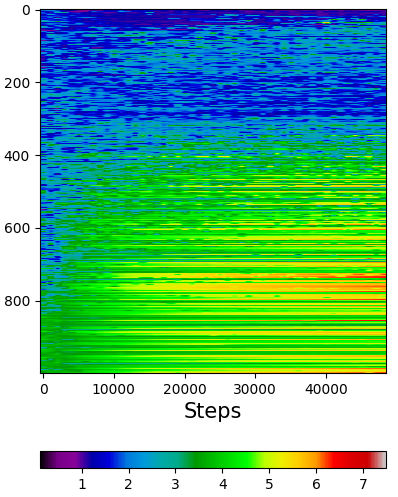}  
    \end{tabular}
\end{table}

\begin{table}[h]
    \centering
    \begin{tabular}{p{0.05cm} p{0.05cm} c c c}
     &  &  $n = 2$  & $n = 3$ & $n = 4$  \\
     \raisebox{7\normalbaselineskip}[0pt][0pt]{\rotatebox[origin=c]{90}{ $H(O,S)$}} &      \raisebox{7\normalbaselineskip}[0pt][0pt]{\rotatebox[origin=c]{90}{\small agents}} &    \includegraphics[width = 4.25cm]{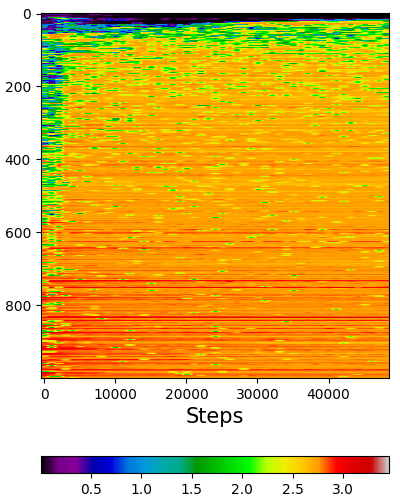}   &         \includegraphics[width = 4.25cm]{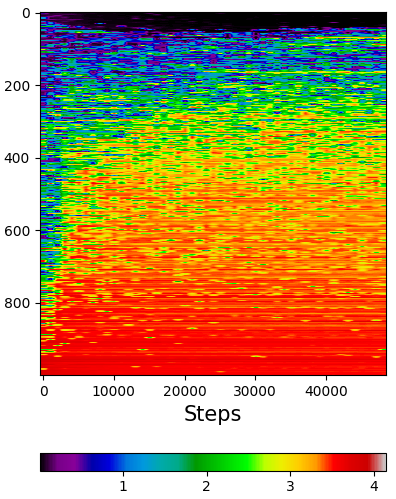} &  
            \includegraphics[width = 4.25cm]{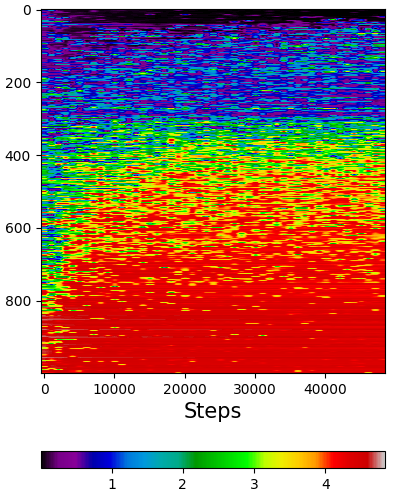}  \\
                 \raisebox{7\normalbaselineskip}[0pt][0pt]{\rotatebox[origin=c]{90}{$H( O)$}} &      \raisebox{7\normalbaselineskip}[0pt][0pt]{\rotatebox[origin=c]{90}{\small agents}} &  
           \includegraphics[width = 4.25cm]{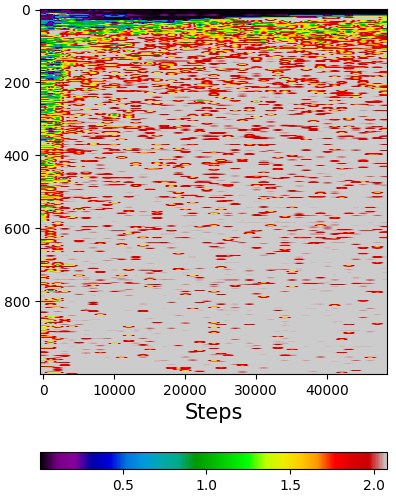}   &         \includegraphics[width = 4.25cm]{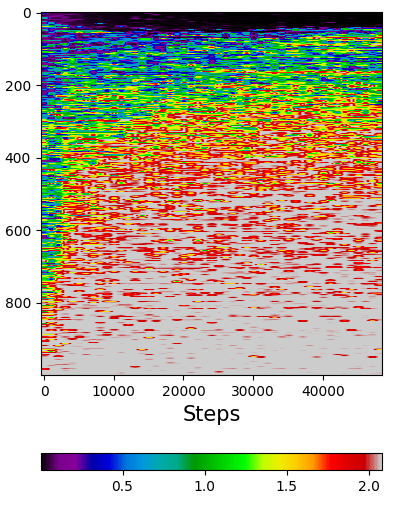} &  
            \includegraphics[width = 4.25cm]{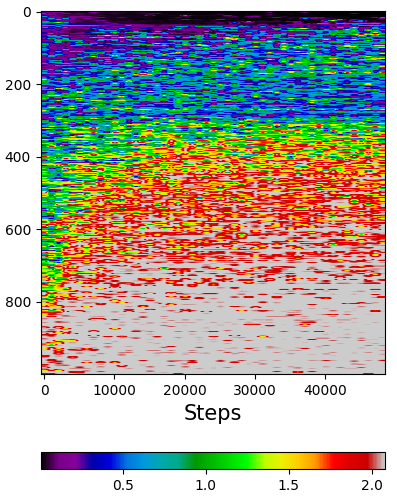}   \\
     \raisebox{7\normalbaselineskip}[0pt][0pt]{\rotatebox[origin=c]{90}{$H( O^{\star})$}} &      \raisebox{7\normalbaselineskip}[0pt][0pt]{\rotatebox[origin=c]{90}{\small agents}} &  
           \includegraphics[width = 4.25cm]{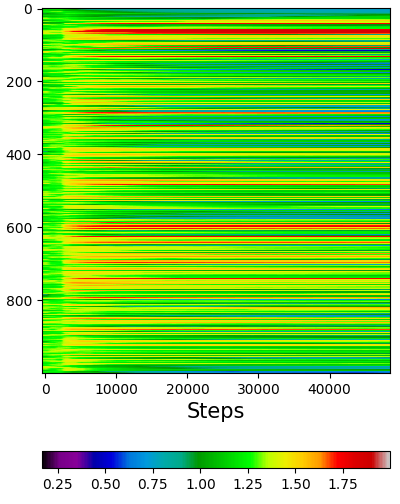}   &         \includegraphics[width = 4.25cm]{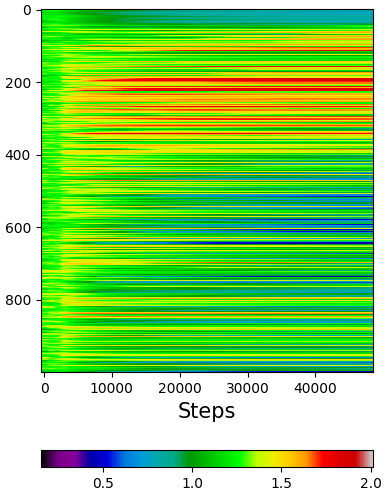} &  
            \includegraphics[width = 4.25cm]{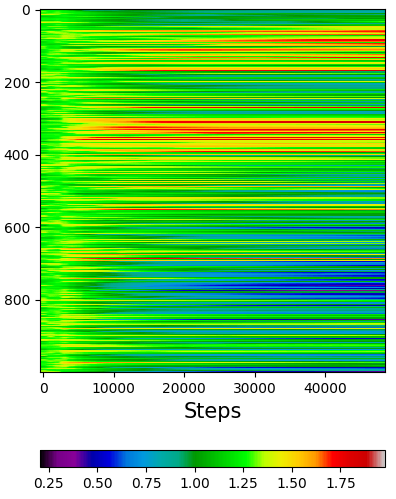} 
    \end{tabular}
\end{table}
\end{document}